**Ultimate parameters of an all-optical $M_X$ resonance in Cs in ultra-weak magnetic field**
Revised 7/10/2023  22:55:00

*M.V. Petrenko[1], A.S. Pazgalev[1], and A.K. Vershovskii[1]*

[1]*Ioffe Institute, Russian Academy of Sciences, St. Petersburg, 194021 Russia*
*e-mail address: antver@mail.ioffe.ru*

We present the results of studying the parameters of the magnetic $M_X$ resonance in an all-optical sensor built according to the two-beam Bell-Bloom scheme in nonzero ultra-weak magnetic fields in which the effects of spin-exchange broadening suppression are partially manifested. We report on the features of the resonance under these conditions. We also optimize the resonance parameters to achieve maximum sensitivity in magnetoencephalographic sensors. We demonstrate an improvement in the ultimate achievable sensitivity of an all-optical $M_X$ sensor by a factor of four or more, which in our experiment corresponds to a decrease from 13 to 3 fT/Hz$^{1/2}$ in a volume of 0.13 cm$^3$. We also report the effect of incomplete suppression of spin-exchange broadening under conditions of strong transverse modulated optical pumping, and propose a semi-empirical model to describe it.

**Keywords:** *Optically detected magnetic resonance, quantum magnetometer, ultra-weak magnetic field, magnetoencephalography.*

## I. INTRODUCTION

This work is devoted to the search for promising ways to develop compact highly sensitive magnetic field (MF) sensors for the tasks of studying the magnetic activity of the brain (magnetoencephalography, MEG). This uniquely informative biomedical method, which appeared more than half a century ago, was not widely used for a long time, since the only type of sensors that provided the necessary sensitivity with sufficient compactness were SQUIDs (Superconducting Quantum Interference Devices) [1]. SQUID sensor arrays are extremely expensive, as is their annual maintenance; in addition, the SQUID sensors are separated from the object (human head) by a thick-walled dewar helmet, which significantly reduces the level of useful signal and limits the spatial resolution of the system.

The main competitor for SQUID appeared with the invention of optically pumped sensors using the SERF (Spin Exchange Relaxation Free) mode, which involves the suppression of spin-exchange broadening in a zero MF [2–6]. This caused an explosion of interest in MEG methods. SERF sensors are based on optical pumping and zero-field magnetic resonance (MR) line detection [2,3,7–10]. They do not require liquid helium cooling; they can be brought much closer to the head than SQUID sensors; they are capable of simultaneously measuring two or even three MF components. However, they also have disadvantages: an MEG array based on SERF is operable only if the field inhomogeneity in the volume does not exceed several tens of nanoteslas, and this nearly excludes the possibility of working in relatively compact magnetic shields. In addition, each sensor modulates the MF in two or three [11] perpendicular directions, as a result of which the sensors operating in the array influence one another.

On the other hand, optically pumped non-zero field sensors (namely, $M_X$ sensors) demonstrate parameters that are somewhat worse, but sufficiently suitable for use in MEG. [12,13]. $M_X$ sensors do not use the SERF effect and measure not the MF components, but its modulus [10,14,15]. It should be noted here that in the presence of a non-zero constant field ***B***, such scalar sensors turn out to be sensitive only to the component of a weak signal that is parallel to ***B***; consequently, in MEG, the $M_X$ sensors turn out to be single-component with the sensitivity axis determined by the direction of the external field. $M_X$ sensors, unlike SERF, can be "quiet", meaning that they not use either MF modulations or any other RF fields; they impose much lower requirements on the homogeneity of the MF; and due to the significantly larger width of the MR, they provide greater speed. In addition, the $M_X$ signal is recorded at the Larmor frequency proportional to the MF, which makes it possible to transfer the signal frequency to a range free from low-frequency technical laser noise. Unfortunately, a wider MR linewidth to achieve the same level of sensitivity requires higher pump speeds – and therefore more powerful lasers, which, as a rule, cannot be integrated directly into the sensor (as is done in sensors using SERF effect [7,16]).

In this paper, we explore the possibility of a compromise solution, namely, the use of an all-optical "quiet" $M_X$ sensor built according to the Bell-Bloom scheme [12,14,17–19] in ultra-weak MFs (meaning fields in which the SERF effect is already manifested, but the Larmor frequency still exceeds the linewidth). We demonstrate that such a sensor can show parameters similar to those of the SERF sensor without using MF modulation. At the same time, the MF required for such a sensor with an induction from one to several hundreds of nanotesla can be created both in a magnetically isolated room and in a magnetic shield, which promises a significant expansion of the field of application of the MEG. We demonstrate an improvement in the ultimate sensitivity (estimated from the MR parameters and shot noise) of the MX sensor in ultra-weak MFs by a factor of four or more, which in our particular scheme corresponds to an improvement from 13 to 3 fT/Hz$^{1/2}$ in a volume of 0.13 cm$^3$.

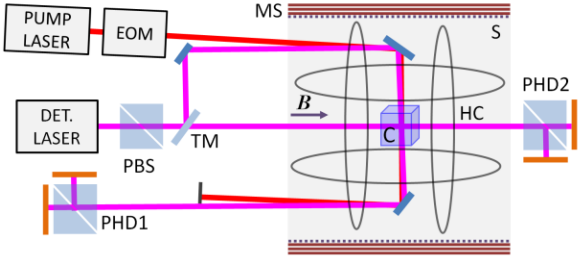

FIG. 1. A simplified experimental setup: PUMP LASER, a pump light source; DET. LASER, detection light source; EOM, electro-optical modulator; M, mirror; TM, translucent mirror; C, gas cell with Cs vapor; MS, magnetic shield; S, solenoid; HC, Helmholtz coils system; PBS, polarization cube; PHD1 and PHD2, balanced photodetectors. The cell C is enclosed in a thermostat (not shown in the diagram).

This article is organized as follows: in Section II, we describe the experimental setup and define the direction of the research. In Section III, we present the theoretical calculations that make it possible to estimate the dependence of the MR width on the MF and pump parameters. In Section IV, we present the data of the MR width measurement experiment and compare them with the data in Section III. In Section V, we examine the achievable ultimate sensitivity. In Section VI, we discuss our findings and present options for explaining the observed effects.

## II. EXPERIMENTAL SETUP

The study of the parameters of the MR was carried out in a magnetic shield [20]. A cell with an internal size of 8×8×8 mm$^3$ filled with saturated Cs vapor and nitrogen at a pressure of 100 torr was used as the sensitive element. Cesium vapors were optically oriented by the pump light emitted by an external cavity diode laser (VitaWave company). The MR signal was detected by the probe light emitted by a similar laser.

The experiment is carried out in the Voigt geometry [20]: both beams, the pump beam and the detection (or probe) beam, were perpendicular to the MF and almost parallel to each other. Their Gaussian diameter was 4.9 mm, so that the pumped/detected volume was 0.13 cm$^3$ (which roughly corresponds to the volume of a cubic cell with a side of 5 mm). Part of the intensity of the detecting beam was split off on a translucent mirror TM and passed through the cell in the direction parallel to the magnetic screen axis (along vector $\boldsymbol{B}$). The pump beam frequency is tuned to resonance with the $D_1$ transitions of the cesium line, which connect the hyperfine level $F = I-\frac{1}{2}$ of the $S_{1/2}$ state with the levels $F' = I\pm\frac{1}{2}$ of the $P_{1/2}$ state [21,22]. The effective Zeeman pumping of the sublevel $F = I+\frac{1}{2}$, $m_F = F$ is partly due to the overlap of the optical absorption contours of two hyperfine transitions, and partly due to the conservation of the nuclear component of the momentum during collisions with buffer gas atoms in an excited state [22]. The MR was excited according to the Bell-Bloom scheme [12]: the polarization of the pump radiation was modulated by means of an EOM at a frequency close to the MR frequency: half a period, left-hand circular polarization; half a period, right-handed. A similar experimental geometry, but with an RF field, was used, for example, in [23].

At each temperature and/or at each MF value, we took an array of records at four values of the pump intensity and at four values of the detection intensity (16 records of the MR signal in total). The signals of rotation of the polarization azimuth of the probe beams were detected by balanced photodetectors PHD1 and PHD2, after which they were fed to SR830 synchronous detectors. The polarization azimuth rotation signal from the PHD2 photodetector was used as an error signal for zeroing the transverse field components, which was carried out using a two-coordinate Helmholtz coils system [24,25]. For further processing, we used the polarization rotation signal from the PHD1 photodetector, which is proportional to the precessing transverse magnetization of the medium. Two quadrature ($S_x$ and $S_y$) components of the signal from the output of the synchronous detector were combined into a complex signal $S_x + iS_y$ and approximated by a common Lorentzian contour with an arbitrary phase, while the complex bias was subtracted from them. Next, the obtained series were processed and the MR parameters were extrapolated to zero detection intensity.

## III. THE THEORY OF RELAXATION IN ULTRA-WEAK FIELDS

The effects that determine the resonance width in ultra-weak magnetic fields are extremely complex; therefore, there is still no unified theory describing them under all conditions and experimental configurations, despite the enormous work done by W. Happer and colleagues in [26–28], by M. Romalis and colleagues [23,29], and subsequently systematized in [30]. Particularly poorly studied are the relaxation processes under conditions of strong modulated transverse pumping, which are becoming increasingly important with the development of "quiet" MEG sensors.

According to [26] and [30] (Eqs. 2.133–2.134), the longitudinal relaxation rate $G_1 = 1/T_1$ is

$$G_1 = \frac{1}{q_N}(R_{SD} + R_{BG} + R_{OP} + R_D) + R_W, \qquad (1)$$

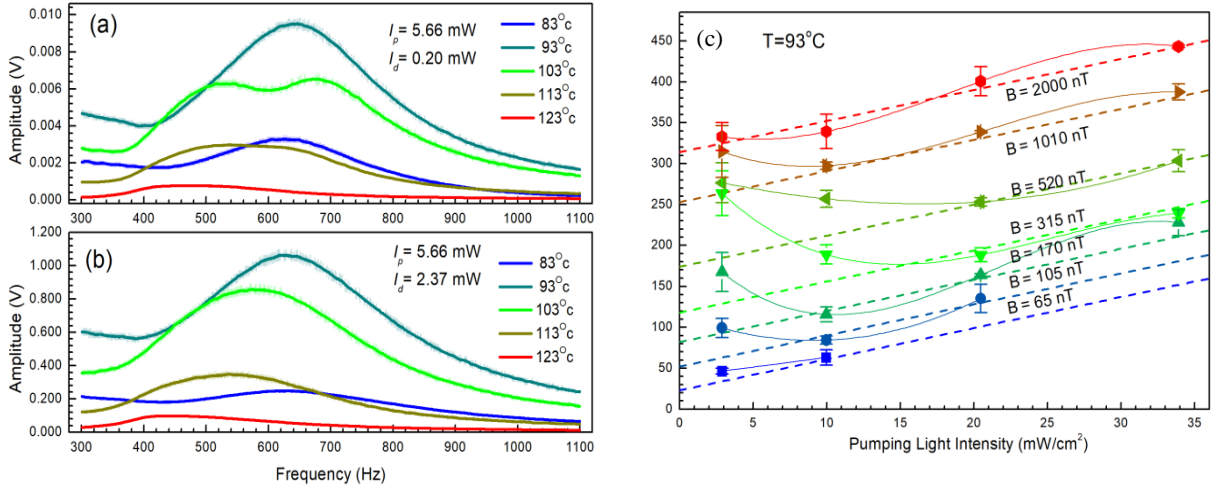

FIG. 2. (a),(b) Examples of MR recordings at $B = 65$ nT. (c) Dependences of the MR width on the pump intensity. Each point was obtained by extrapolating the power of the detecting radiation to zero. Dashed lines are the result of extrapolation of the linear part of the data sets to zero pump intensity; they are all assumed to have the same slope. Solid lines are approximating polynomials.

where $R_{SD}$ is the spin destruction rate in collisions of alkali atoms with each other, $R_{BG}$ is the destruction rate in collisions with a buffer gas, $R_{OP}$ is the optical pumping rate, $R_D$ is the rate of destruction by detecting light, $R_W$ is the rate of destruction by collisions with walls; $q_N$ is the nuclear "slowing-down factor" equal to the ratio of the total moment of the atom to the electronic part of the moment, and dependent on the degree of polarization of the medium P: in the case of Cs, $q_{N0} = 22$ and $q_{N\infty} = 2I+1 = 8$ (hereinafter, index 0 corresponds to $R_{OP} = 0$, the index $\infty$ corresponds to $R_{OP} \to \infty$); $I$ is the atomic nuclear momentum.

Below, we will consider the resonance widths extrapolated to zero intensity of the detecting light ($R_D \to 0$). The dark (i.e., extrapolated to zero pump intensity) longitudinal relaxation rate of the atomic moment is

$$R_{10} = R_{SD} + R_{BG} + R_W, \quad (2)$$

and the corresponding dark width of the MR is

$$G_{10} = \frac{1}{q_N}(R_{SD} + R_{BG}) + R_W. \quad (3)$$

According to [30] (Eq. 2.39), in the first approximation

$$P \approx \frac{R_{OP}}{R_{10} + R_{OP}}. \quad (4)$$

The transverse relaxation rate $G_2 = 1/T_2$ is

$$G_2 = G_{10} + G_{OP} + G_{SE}, \quad (5)$$

where $G_{OP} = R_{OP}/q_N$ is the effective pump rate, $G_{SE} = R_{SE}/q_{SE}$ is the spin-exchange broadening of the MR line, $R_{SE}$ is the spin-exchange relaxation rate, and $q_{SE}$ is the coefficient of suppression of spin-exchange broadening. In ultra-weak fields $q_{SE}$ tends to infinity (SERF effect), and in the strong field limit ( [27], also see [30] (Eq. 2.138)) it is equal to

$$q_{SEH} = \frac{3(2I+1)^2}{2I(2I-1)} \quad (6)$$

(the index $H$ here and below will denote the limit of a strong MF). In almost the entire range of operating temperatures and pumping intensities, the direct contribution of $G_{10}$ to Eq. (5) can be neglected in comparison with $G_{SE}$ and $G_{OP}$; nevertheless, $G_{10}$ plays an essential role in relaxation processes.

As the MF decreases to $B \approx R_{SE}/\gamma$, the SERF effect begins to manifest itself as a decrease in the MR width $G_{SE}$ and the MR frequency $\omega_q$. According to [27] (Eq. 99) and [30] (Eq. 5.5), $G_{SE0} = \text{Re}[\omega_{SE0}]$, $\omega_{q0} = \text{Im}[\omega_{SE0}]$, where

$$\frac{\omega_{SE0}}{R_{SE}} = M - \sqrt{-\left(\frac{\omega_L}{R_{SE}}\right)^2 - \frac{2i}{[I]}\left(\frac{\omega_L}{R_{SE}}\right) + M^2}, \quad (7)$$

$\omega_L = \gamma B$, $[I] = 2I + 1$, $M = ([I]^2 + 2)/(3[I])$. Here we have rewritten this formula in a simpler form compared to [27] by normalizing frequencies to $R_{SE}$.

The presence of optical pumping leads to three effects: first, according to (5), it causes a linear broadening of the resonance; second, according to (3), as $P$ increases, it causes an increase in $G_1$ due to a decrease in $q_N$; third, according to [28], at $P \approx 1$, it leads to light-narrowing, i.e., to a decrease in the spin-exchange broadening due to the formation of a "stretched" state. In ( [30] (Eq. 4.9)) an approximate formula is given, obtained by expanding [30] (Eq. 4.4–4.8) into a series. It describes the light-narrowing under strong longitudinal pumping and an extremely weak radio frequency field:

$$G_{SE} \approx \frac{G_{SD}}{G_{OP}} \cdot f(\omega_L, R_{SE}) \cdot G_{SE0}, \quad (8)$$

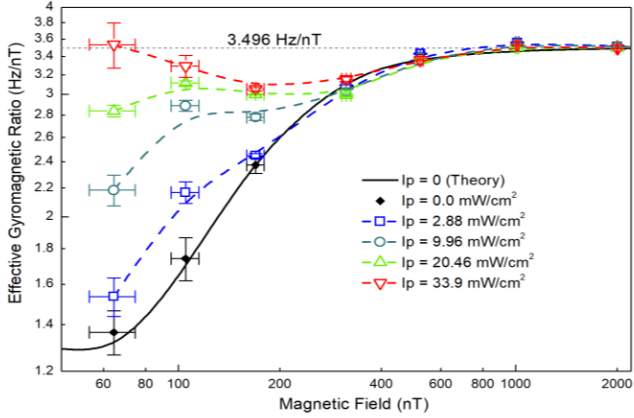

FIG. 3. Dependence of the effective gyromagnetic ratio on the MF induction for different pumping intensities: dots: experiment; solid line: calculation by formula (7); dotted lines: polynomial approximation.

$f(\omega_L, \infty) = 0$. According to Eq. (8), the spin-exchange is completely suppressed under strong pumping, which is also confirmed by the numerical calculation according to [30] (Eqs. 4.4-4.8).

## IV. EXPERIMENTAL INVESTIGATION OF THE RESONANCE WIDTH

We have studied the MR parameters in the range of ultra-weak ($\omega_L = (0.13 – 4.0)\ R_{SE}$) MFs at various temperatures, pump and detection intensities. Examples of MR records at $B = 65$ nT are shown in FIG. 2 (a,b). It can be seen that in this weak field the MR cannot be described by a Lorentzian contour for all experimental parameters; this significantly complicates the interpretation of the data and leads to an increase in errors. This is especially noticeable at low detection intensities and high temperatures, at which nonlinear absorption effects are especially pronounced (FIG. 2(a), T ≥ 103°C). As shown in [23], the reason for the distortion of the shape of resonances in the case of incomplete coincidence of the pump and interrogation zones can be atomic diffusion.

FIG. 2(c) shows the experimentally measured dependences of the MR width on the pump intensity: each point is the result of extrapolation of the MR width to zero probe intensity over a series of measurements. Error values were obtained from the extrapolation dispersion.

From the data FIG. 2(c), the $G_{SE0}$ values were obtained by polynomial approximation, and the $G_{SE\infty}$ values were obtained by extrapolation of the linear part of the data sets to zero pump intensity. With linear extrapolation in accordance with Eq. (5), it was assumed that the slope of all straight lines is the same; it amounted to $k_p = 2\pi\ (3.8 \pm 0.3)$ Hz·mW$^{-1}$·cm$^2$. At the same time, it turns out (FIG. 6) that for all studied MF values, as well as in a wide temperature range, the $G_{SE0}$ value (as expected) is described by Eq. (7), while the $G_{SE\infty}$ value demonstrates a new unexpected dependence.

## V. EXPERIMENTAL ESTIMATION OF THE ULTIMATE SENSITIVITY

The ultimate sensitivity of the sensor can be estimated [31] based on the MR parameters and noise spectral density:

$$\delta B \approx \frac{1}{\gamma}\frac{G_2 \cdot \rho_N}{A}, \qquad (9)$$

where A is the MR amplitude. Photon shot noise, whose value we use for further estimates, is not the only fundamental noise. Strictly speaking, the limit is the sensitivity, limited by the sum of photon and atomic projection noise. However, as shown in [32], the correct optimization of the sensor parameters leads to an approximate equality of the spectral densities of atomic and photon noise, and taking into account only one of them leads to an overestimation of the sensitivity by a maximum of one and a half to two times. The validity of this statement for various schemes of the $M_X$ sensor was demonstrated by us in [20,33].

As it was shown above (FIG. 6), as the MF decreases, the resonance width decreases by an order of magnitude or more. The amplitude of the resonance in this case first slightly increases, then begins to fall. But Eq. (9) includes one more parameter – namely, the gyromagnetic ratio $\gamma$, which is considered a constant in strong magnetic fields ($\gamma = \gamma_H$), and which is not constant in ultra-weak fields. FIG. 3 shows the theoretical dependence $\gamma(B)$ constructed from Eq. (7) and the values of $\gamma$ measured at different pump intensities. As we can see, in the absence of pumping, the "dark" value of $\gamma_0$ decreases with decreasing MF. The $G_2/\gamma_0$ ratio remains approximately constant up to $\gamma_H B \approx 0.3\ R_{SE}$, and only then begins to decrease. This could deprive us of any hope of improving the sensitivity, but, fortunately, the shift of the MR line by the pump light maintains the effective value of $\gamma$ at a level close to $\gamma_H$.

We have studied the dependence of the ultimate sensitivity on the pump and detection intensities in a wide range of MFs and temperatures. FIG. 4 shows data arrays obtained at $B = 105$ nT. It follows from the figure that the sensitivity optimum exists for both intensities and for temperature, and with decreasing temperature, the required intensities decrease (primarily, the pump intensity). FIG. 5 shows graphs of the ultimate sensitivity depending on the MF and temperature, obtained at each point by a sample from an array consisting of 16 measurements.

The graphs show a more than fourfold improvement in sensitivity in an ultra-weak MF, from $(13.2 \pm 0.2)$ fT/Hz$^{1/2}$ to $(2.96 \pm 0.17)$ fT/Hz$^{1/2}$. In this case, a decrease in the MF from values $B > 1000$ nT to $B \approx 100$ nT leads to a decrease in the required pump intensity by about a factor of 1.5 and detection intensity by a factor of three, which should be considered an additional advantage. The signal width at the optimum point is $(129 \pm 2)$ Hz.

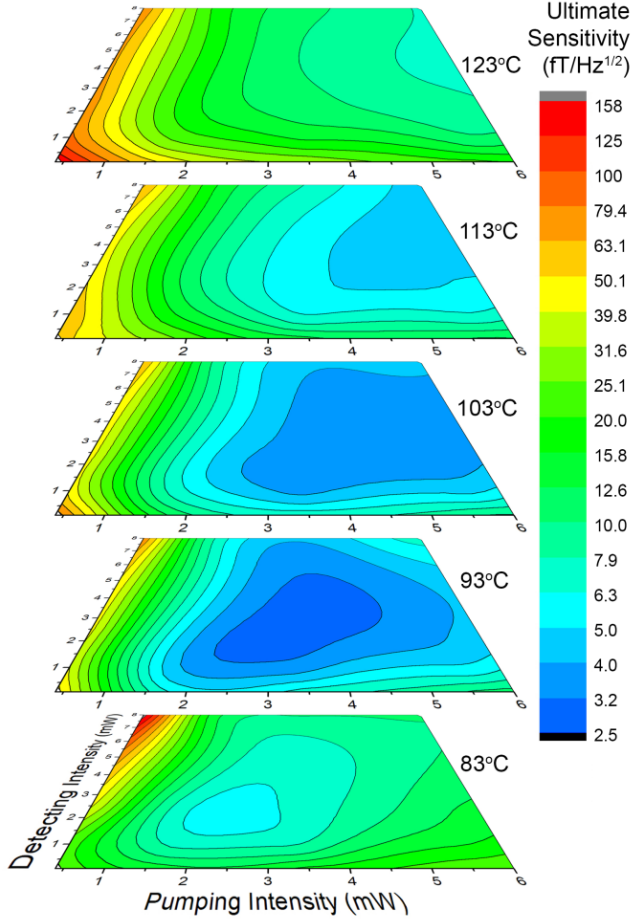

FIG. 4. Dependence of the ultimate sensitivity, estimated from the ratio of the experimentally studied resonance steepness to the calculated spectral density of shot noise, on the pump and detecting intensities and temperature.

## VI. DISCUSSION

The most unexpected result, following from the above experimental data, is that in the case of transversely modulated pumping, the spin-exchange width $G_{SE}$ tends to a certain nonzero value $G_{SE\infty}$ with increasing intensity. The difference between our experiment and [28] is that, instead of a weak RF field, we used modulation of the ellipticity of the pump light. This also distinguishes our experimental scheme from numerous SERF schemes (see, for example, [23]), which, like ours, use transverse pumping. Thus, the presence of $G_{SE\infty} > 0$ is due to the modulation of the pump light, and not to its direction.

The narrowing of the MR line in Cs was also studied in [34] in the scheme of an RF magnetometer with longitudinal pumping; from the data presented by the authors, it can be concluded that in their experiment the residual broadening is also not equal to zero. But since the authors do not provide data for different MF values, it is difficult to say whether this broadening is due to spin exchange or the contribution of longitudinal relaxation.

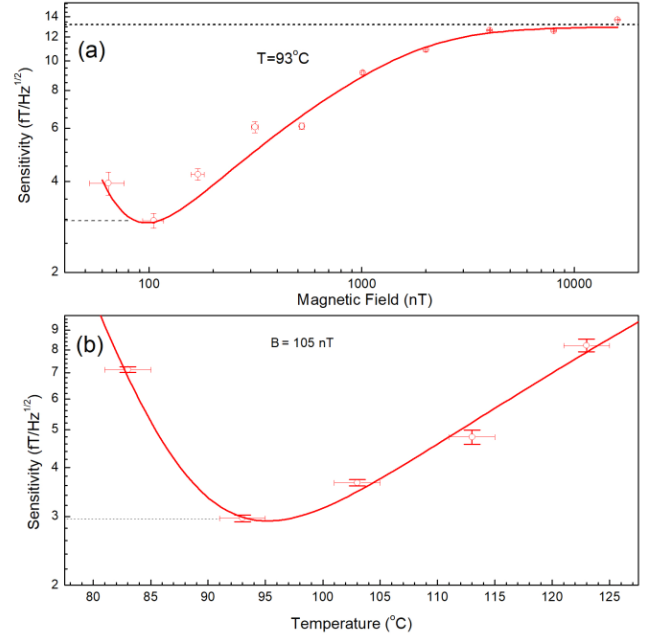

FIG. 5. Dependence of the ultimate sensitivity, estimated from the ratio of the experimentally studied resonance steepness to the calculated spectral density of shot noise (a) on the magnetic field, (b) on temperature. The dots are the experimental results, the lines are guides for the eye.

In our previous works [35] we have already noted the fact that the narrowing of the line by light in strong MF turns out to be much weaker in the Bell-Bloom scheme than in the "classical" scheme of the $M_X$ sensor [36]. We believe that the physical nature of this limit is related to the features of the pump modulation.

The modulation of the pump ellipticity at a constant intensity should be rectangular with a constant (namely, the maximum possible) amplitude; amplitude reduction is unacceptable, because it is equivalent to mixing linear polarization with circular polarization. Therefore, the angular momentum introduced by the pump light turns out to be coaxial with the precessing atomic momentum twice per period, while in other modulation phases it makes an angle with it from $-\pi/2$ to $\pi/2$. Consequently, any additional phase delay of the atomic moment will lead to the fact that part of the pump radiation will not pump, but, on the contrary, will dephase the collective moment. Under conditions close to the SERF regime, the dynamics of the system are determined by the spin-exchange rate $R_{SE}$: any change in external parameters (for example, the polarization of pump light) leads to the establishment of a new equilibrium state in the system with a characteristic time $\Delta t = 1/R_{SE}$, and to a phase delay $\Delta \varphi = \omega_L/R_{SE}$. Therefore, by analogy with Eq. (4), we can introduce the parameter of phase coherence, or phase polarization:

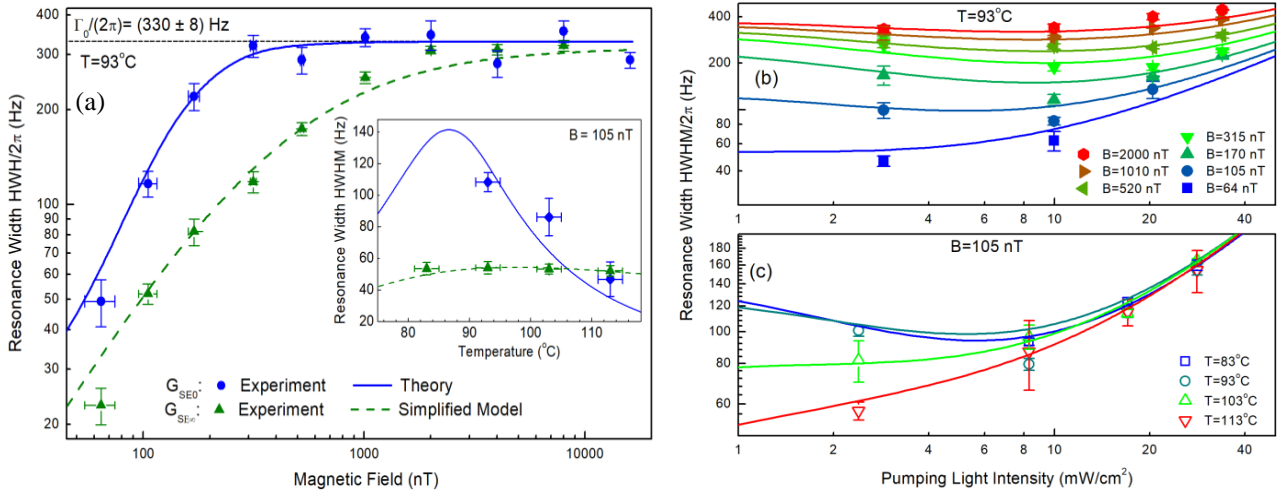

FIG. 6. (a) Dependence of the dark ($G_{SE0}$) and maximally light-reduced ($G_{SE\infty}$) spin-exchange relaxation rates as functions of the MF; points: experiment, lines: calculation by Eqs. (7) and (12). The inset shows the same values depending on the temperature. (b) Dependences of the magnetic resonance width on the pump intensity at $T = 93°C$ for several values of the MF induction; solid lines: calculation by Eq. (12). (c) The same values at $B = 105$ nT for several temperatures.

$$P_\varphi \approx \frac{R_{SE}}{R_{SE} + \omega_L} \quad (10)$$

Let us try to build a simple semi-empirical model for the narrowing of the MR line by light. The rate of residual spin exchange $G_{SE}$ upon narrowing of the line by light is determined by the population of the levels adjacent to the level being pumped. In the first approximation, this population is proportional to $(1-P)$:

$$G_{SE} = (1-P)G_{SE0} \quad (11)$$

Note that, under strong pumping, $(1-P) = G_1/G_{OP} \approx G_{OP}/G_{SD}$, and Eq. (11) goes over into Eq. (8) given earlier. Let us also take into account in Eq. (11) the possible dephasing of the atomic magnetic moment by pump modulation. The MR broadening $G_M$ introduced by the modulation should be zero in the absence of pumping, and constant with strong pumping. Therefore, we can assume that $G_M$ is proportional to $P$. As follows from FIG. 6(a), the value of $G_M$ in weak fields increases in proportion to the modulation frequency and saturates at the level $G_{SEH} = R_{SEH}/q_{SEH}$ at $\omega_L \approx R_{SEH}$. Therefore, we can also assume that $G_M \sim (1-P_\varphi)$:

$$G_{SE} = (1-P^n)G_{SE0} + P(1-P_\varphi)^m G_{SEH} \quad (12)$$

Empirical parameters $n$ and $m$ are introduced here in order to more adequately describe a multilevel system, for which the polarization coefficients are determined differently than for a two-level one.

The result of the approximation of the experimental data by Eqs. (7),(12) at $n = 1.95$, $m = 1.1$ is shown in FIG. 6. It should be noted that the data obtained in an MF with an induction of more than 2000 nT reveal other effects, in particular, an increase in the coefficient $k_p$. These effects are outside the scope of this article and will not be discussed here.

An example of the dependence of the total resonance width $G_2$ and individual contributions to the width on the pumping rate calculated in our model is shown in FIG. 7; in addition, FIG. 7 shows the dependence of the total width $G_{WH}$, calculated by the formulas Eqs. 4.4–4.8 from [30].

It follows from Eqs. (7), (12) that under transverse modulated pumping, the spin-exchange broadening in weak fields can be reduced by a maximum of a factor of three by pumping light. On the other hand, according to [30], the total resonance width (which includes broadening by pumping light $G_{OP}$), in Cs can be reduced by the longitudinal pump light by a factor of eight. But as one can see, it decreases by no more than a factor of two in the Bell–Bloom configuration.

As we suggested above, the residual broadening by modulation can be due to two factors: 1) the width of the distribution of the phase of the modulation itself (from $-\pi/2$ to $+\pi/2$ with respect to the phase of the fundamental harmonic of the modulation, and 2) the phase delay of the atomic moment with respect to the fundamental harmonic of the modulation. Note that the most obvious assumption that the side spectral components of the modulated radiation are the broadening factor does not work, since these components lie far beyond the width of the MR line. In order to find out which of the two factors is decisive, we carried out an additional experiment, replacing the modulation of the pump ellipticity by the modulation of its intensity. In this case, pumping was carried out once per period of the Larmor frequency by a pulse with constant circular polarization and variable duration (from 10 to 50%). The resonance width was measured as a function of the pulse duration at a constant average pump intensity. At $B = 105$ nT, the maximum broadening by a long pulse was $(13.0 \pm 7.5)$ Hz, which is an insignificant part of the residual broadening $G_{SE\infty} = (52 \pm 4)$ Hz.

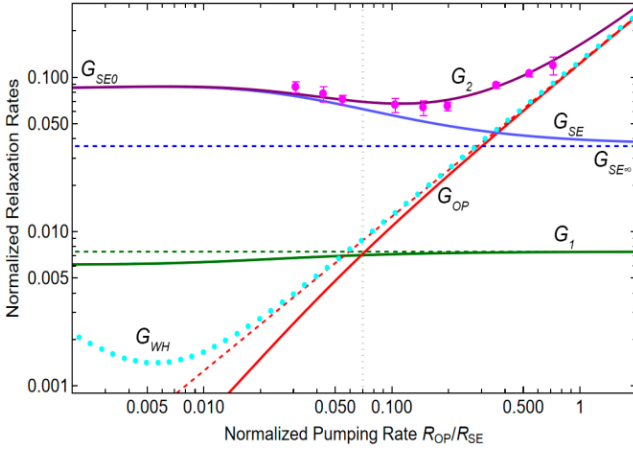

FIG. 7. An example of the dependence of the calculated (by Eqs. (7)-(12)) MR width on the effective pumping rate (black line) and its comparison with experiment (magenta dots). $R_{WH}$ – calculation by formulas Eq. 4.4-4.8 from [30]; other designations are given in the text. All quantities are normalized to the spin exchange rate $R_{SE}$. Calculation parameters: $T = 93^\circ$C, $B = 105$ nT, $\omega_L = (2\pi)367$ Hz. Calculated relaxation rates: $R_{SE} = (2\pi)1438$ Hz, $G_{SE0} = (2\pi)119$ Hz, $G_{SE\infty} = (2\pi)52$ Hz, $G_W = (2\pi)7.7$ Hz, $R_{SD} = (2\pi)10.3$ Hz, $R_{BG} = (2\pi)13.3$ Hz.

At $B = 2000$ nT, the maximum broadening by a long pulse turned out to be indistinguishable within the error. Thus, we have shown that factor #1 is not decisive, and the residual broadening is caused by the phase shift of the momentum precession with respect to the pump.

Even taking into account the incomplete light narrowing of the MR line, the results presented above demonstrate a significant advantage of ultra-weak MFs for the operation of an $M_X$ sensor built according to the Bell-Bloom scheme. However, there are two negative factors that should be taken into account: firstly, a decrease in line width means a decrease in instrument performance; secondly, lowering the MR frequency to hundreds of hertz makes the sensor potentially more sensitive to low-frequency technical noises of the laser, and, accordingly, requires special measures to suppress these noises (signal frequency transfer by modulation of the detecting light can be used as one such measure [37–39]). In other words, when reducing the width of the MP line and the polling rate in MX sensors, the same problems arise as in SERF sensors. The balanced recording scheme used in our experiment is capable of suppressing the laser intensity noise by one to two orders of magnitude compared to simple recording of the transmitted light intensity used in SERF compact sensors, but this is achieved at the expense of the scheme complexity.

## VII. CONCLUSIONS

We have demonstrated the fundamental performance of the Bell-Bloom $M_X$ sensor in ultra-weak magnetic fields and showed that its sensitivity can be improved to reach that of a SERF sensor, while retaining most of the advantages and disadvantages of the $M_X$ scheme. Among these advantages is the absence of radio frequency interference, among the disadvantages is the inability to simultaneously measure more than one component of the MEG signal. In addition, in ultra-weak fields, the speed of the sensor decreases and the level of technical noise increases, which requires special precautions. On the other hand, in ultra-weak field, the $M_X$ sensor requires substantially lower intensities for pumping and detection.

We also showed that spin-exchange relaxation under conditions of transversely modulated pumping has its own peculiarities and gave an empirical description of them.

## VIII. CONFLICT OF INTEREST.

The authors declare that they have no conflict of interest.